\RequirePackage{fix-cm}
\documentclass[pdftex,twocolumn,epjc3]{svjour3}          

\RequirePackage[T1]{fontenc}

\smartqed  

\RequirePackage{graphicx}
\RequirePackage{amssymb}
\RequirePackage{mathptmx} 
\RequirePackage{fdsymbol}
\RequirePackage{flushend}
\RequirePackage[numbers,sort&compress]{natbib}

\RequirePackage[colorlinks,citecolor=blue,urlcolor=blue,linkcolor=blue]{hyperref}
\RequirePackage{xcolor}
\usepackage[switch]{lineno}
\usepackage{multirow}
\usepackage[separate-uncertainty=true]{siunitx}

\journalname{Eur. Phys. J. C}
\title{Photoluminescence response of acrylic (PMMA) and polytetrafluoroethylene (PTFE) to ultraviolet light}
\subtitle{Limits on low-intensity photoluminescence in support materials of rare-event search experiments}

\begin{document}

\author{G.R. Araujo \thanksref{e1,addr1}
        \and
        T. Pollmann\thanksref{e2,addr1} \and
        A. Ulrich\thanksref{addr1}
}

\thankstext{e1}{e-mail: garaujo.rodrigues@tum.de}
\thankstext{e2}{e-mail: tina.pollmann@tum.de}

\institute{Physik-Department E15, Technische Universit\"at M\"unchen, James-Franck-Str., 85748 Garching, Germany\label{addr1}
}


\maketitle

\begin{abstract}

Some publications indicate that poly(methyl metha\-crylate) (PMMA) and  poly\-tetra\-fluoroethylene (PTFE) exhibit low levels of photoluminesence (fluorescence and/or phosphorescence) when irradiated with photons in the ultraviolet (UV) to visible range. PMMA (also known as acrylic) and PTFE are commonly used to contain the liquid argon (LAr) or xenon (LXe) target material in rare-event search experiments. LAr and LXe scintillate in the vacuum UV region, and the PMMA and PTFE can be directly illuminated by these photons. Photoluminescence from support materials could cause unexpected signals in these detectors.

We investigate photoluminesence in the \SIrange{400}{550}{\nano\meter} region in response to excitation with UV light between \SI{130}{nm} and \SI{250}{nm} at levels relevant to rare-event search experiments.  Measurements are done at room temperature and the signal intensity is time-integrated over several minutes. 

We tested PMMA and PTFE samples from the batches used in the DEAP-3600 and LUX experiments and observed no photoluminescence signal. We put limits on the efficiency of the plastics to shift UV photons to a wavelengths region of \SIrange{400}{550}{\nano\meter} at 0.05\% to 0.35\% relative to the wavelength shifting efficiency of tetraphenyl-butadiene.

\end{abstract}

\section{Introduction}
\label{sec:intro}
The scintillation and/or ionisation signal of liquid noble gases, particularly liquid argon and liquid xenon, is commonly used in particle detectors looking for rare events  \citep{DarkSide, DEAPconst, protodune, miniclean, microboone, LUX, xenonT, ArDM}. The scintillation signal is in the vacuum ultraviolet (VUV) regime at wavelengths of approximately \SI{128}{\nano\meter} (argon) and \SI{178}{\nano\meter} (xenon) \citep{LArspectra, Doke}.

Some materials not usually considered as sources of photons in such detectors might emit light when excited by UV photons. This photoluminescence is classified as either fluorescence 
or phosphorescence depending on the types of the excited states (singlet and triplet), which result in different lifetimes \citep{fluorescence_book}. Often, it is unclear whether a photoluminescence signal is from a singlet or a triplet state; in that case, the term fluorescence is often used for photons emitted from either state. Unexpected fluorescence from plastic support materials, such as containers or substrates, has long been a concern in industrial applications, such as in chip fabrication for biomedical devices \citep{I,II,III,IV,VI,VII,VIII} and has also become a concern in particle detectors looking for weak signals. Several dark matter and neutrino detectors use or plan to use polymethyl methacrylate (PMMA) \citep{DEAPconst, acrylJUNO, SNO}, commonly known as acrylic, or polytetrafluoroethylene (PTFE) \citep{LUX, DarkSide, xenonT, ArDM}, also known as Teflon\footnote{Teflon is a brand name of The Chemours Company.}, both of which have been claimed to fluoresce or phosphoresce at a low level \citep{I,II,III,IV,VI,VII,VIII,V, acryl_phosph_T, acryl_luminesc_cherenkov, PTFE_fluor,PTFE_fluor2, PTFE_fluor4, PTFE_fluor5, PTFE_phosph, PTFE_long_photolum}.


Any photoluminesence response from these plastics is unlikely to be from the bulk molecules; impurities and defects introduced (on purpose or by accident) in the manufacturing process and handling are expected to play a significant role \citep{II, VIII, PTFE_fluor5}. Therefore, we test samples from the specific batches of PMMA and PTFE used in the DEAP-3600 and LUX dark matter search experiments.





\begin{figure}[htp] 
                \begin{center}
                \includegraphics[scale=0.22]{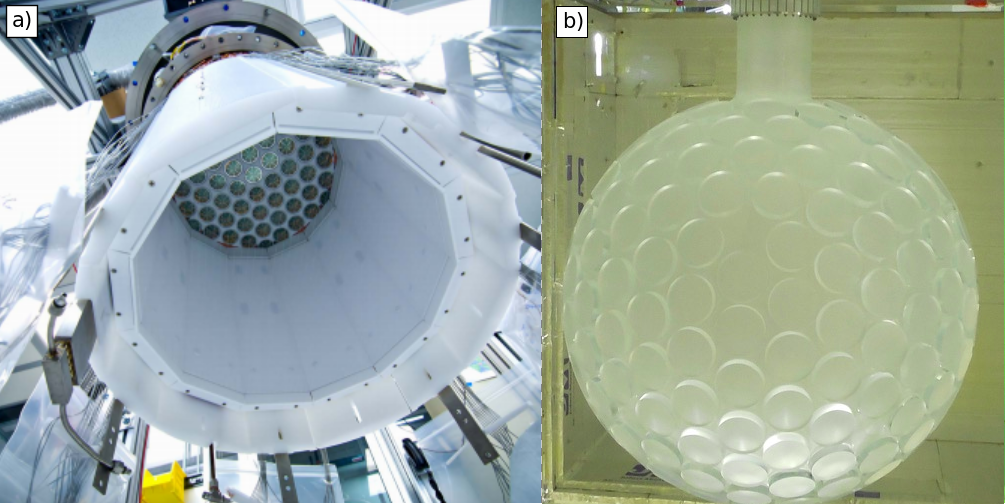}
                \caption{a) View inside of the LUX detector during construction, with
the lower PMT array removed. The cylindrical volume is formed by twelve PTFE panels \citep{LUX}. During operation, the volume is filled with LXe. b) Spherical acrylic cryostat ($\varnothing$\SI{170}{\cm}) used in the DEAP-3600 detector, shown here during construction \citep{DEAPconst}. During operation, the cryostat is filled with LAr.} 
                \label{AV}       
                \end{center}    
        \end{figure}

The liquid xenon (LXe) target in LUX is contained in a cryostat lined with PTFE, as shown in \autoref{AV}a \citep{LUX}. The PTFE used is type 8764 by Technetics \citep{PTFELux}. The LXe scintillation photons are detected directly by VUV-sensitive photomultiplier tubes (PMTs) immersed in the LXe. The PTFE is used to reflect photons impinging on the detector walls until they hit a PMT. If instead of being reflected, photons are absorbed and re-emitted at a later time, the time signature of an event can be distorted \citep{lateelectrontrain}.

The liquid argon (LAr) target in DEAP is contained in a spherical acrylic cryostat shown in \autoref{AV}b. The top of the acrylic sphere is open to a cylindrical acrylic neck which is not coated with TPB. The acrylic is of type UVA and was custom produced by Reynolds Polymer Technologies (RPT). All inner surfaces were sanded, and the spherical region was coated with a thin layer of the wavelength shifter (WLS) 1,1,4,4-tetraphenyl-1,3-butadiene (TPB)~\citep{TPBDEAP}. The fluorescence of TPB is used to shift the LAr VUV scintillation photons into the blue spectral region (\SIrange{400}{550}{\nano\meter}). The photons can then be transmitted through the acrylic vessel where they are detected by Hamamatsu R5912-HQE PMTs. The PMTs are optically coupled to the acrylic vessel by light guides made of acrylic produced by Spartech. Only photons of wavelengths where acrylic is transparent can be detected.

\begin{figure}[htp] 
\begin{center}
    \includegraphics[scale=0.27]{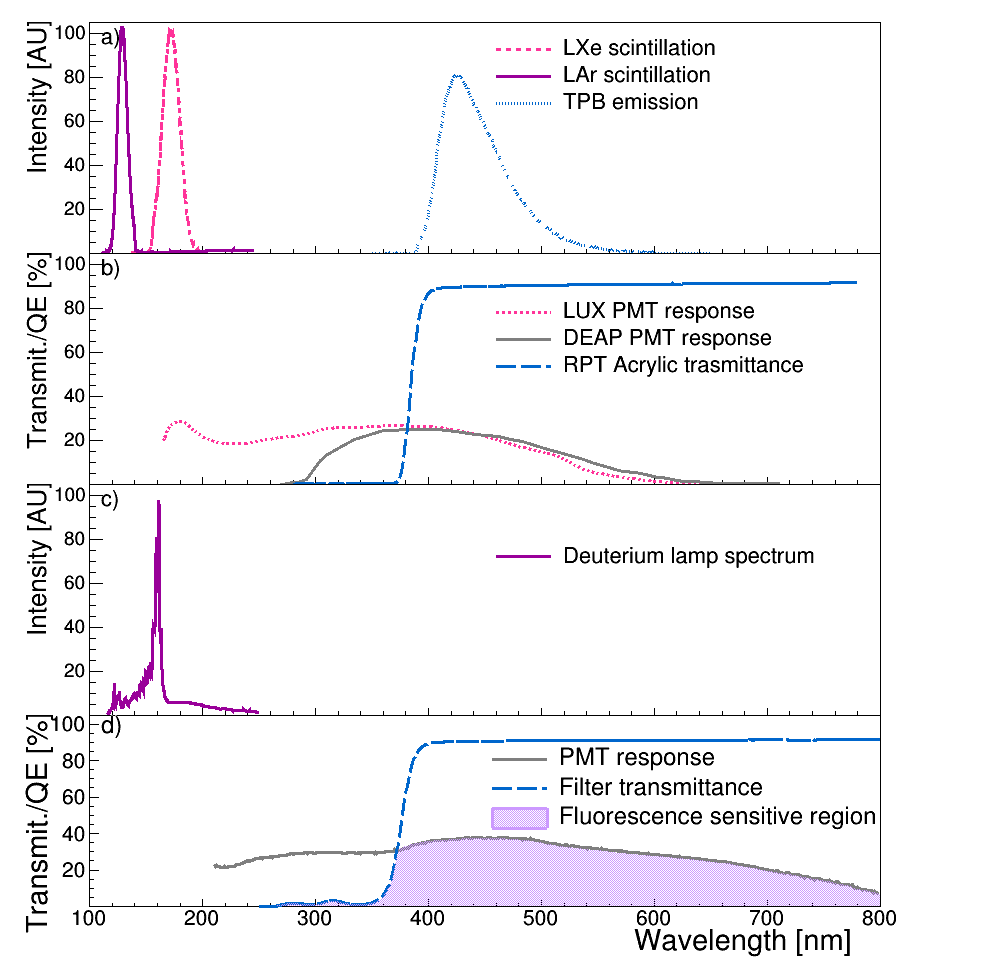}
    \caption{a) Emission spectra of LXe \citep{LXe}, LAr \citep{LArspectra}, and TPB (this work). b) Quantum efficiencies of the Hamamatsu R8778 PMT used in LUX \citep{PMTxenon} and the Hamamatsu R5912 PMT used in DEAP-3600 \citep{r5912_datasheet}. Also shown is the transmittance of a \SI{110}{\mm} thick sample of DEAP acrylic. c) Spectrum of a deuterium lamp. d) Response of the PMT used in this work (combined with the optical fiber) and transmittance of the acrylic filter. The transmittance of the lens was considered to be flat above \SI{200}{\nm} and is not shown here. The shaded area indicates the wavelength region our photon detection setup is sensitive to.}
    \label{LArTPBresponse}  
\end{center}    
\end{figure}

\autoref{LArTPBresponse}a shows the emission spectra of LAr, LXe, and TPB. Also shown is the transmittance of the DEAP acrylic\footnote{Private communication with Victor Golovko, Canadian Nuclear Laboratories.}, and the response of the PMTs used in DEAP and LUX (\autoref{LArTPBresponse}b).

Another source of UV photons in DEAP is Cherenkov light, which can be created inside acrylic by fast electrons from the decay of radioactive contaminants. The Cherenkov spectrum has a UV component which is predicted to be more intense than the visible one \citep{acryl_luminesc_cherenkov} but is expected to be absorbed inside the acrylic. Cherenkov signals could be enhanced by any fluorescence of the acrylic, which would shift the UV photons into a region where acrylic is transparent. Such an enhancement of the Cherenkov photon yield would lead to discrepancies between Monte Carlo simulation and real data.

As an estimate of the lowest level of photoluminescence still relevant for rare-event search experiments using LAr, TPB, and acrylic, we consider the wavelength shifting efficiency (WLSE) PMMA would need to produce events in the energy region of interest for dark matter search, approximately \SI{100}{\kilo\electronvolt}: The highest energy interactions expected in the detector are alpha decays from the radon chain, at energies of \SI{4.5}{MeV} to \SI{8}{MeV}. The LAr scintillation photons from such a decay are nominally shifted by the TPB to wavelengths where acrylic is transparent, with WLSE $\eta_\text{TPB}$. If all the LAr VUV scintillation from an alpha decay were instead absorbed and wavelength-shifted by acrylic with efficiency $\eta_\text{PMMA}$, the event would be reconstructed at energy $E_{\alpha} \cdot \frac{\eta_\text{PMMA}}{\eta_\text{TPB}}$. For an \SI{8}{MeV} alpha decay to reconstruct as a \SI{0.1}{\MeV} event, the acrylic must have a WLSE  $\eta_\text{PMMA} = \frac{0.1}{8} \eta_\text{TPB}$\footnote{This consideration neglects differences in the photon detection probability for photons emitted from different parts of the detector, as well as quenching factors for alpha particles and nuclear recoils.}.


We use a VUV light source and illuminate both TPB and the plastic samples under the same conditions. The light observed in the measurements of the plastic samples is compared to the amount of light re-emitted by TPB. TPB is commonly used in rare-event search particle detectors \citep{DarkSide, TPBDEAP, microboone, miniclean, protodune, ArDM}. By using TPB as a reference sample, the results are directly applicable to many experiments. The wavelength shifting efficiency of TPB as a function of film thickness has been characterized \citep{TPBGehman2, TPBDavis}, so the results can also be applied to other experiments.
 

\section{Samples}
\label{sec:samples}

Photoluminescence observed from a plastic is not necessarily due to the bulk material itself, but can be caused by impurities, defects or additives in the bulk material, or surface contamination \citep{II,VIII, PTFE_fluor5}. Since the UV light does not penetrate far into the material, surface contamination can have a significant influence on the fluorescence measurement. Many common contaminants, such as oil from fingerprints, can fluoresce \citep{skinfluor}. We therefore use samples made from the same batches of material used in the detectors. 

We measured two samples of PMMA from DEAP and one sample of PTFE from LUX.  

\begin{figure}[htp] 
                \begin{center}
                \includegraphics[scale=0.3]{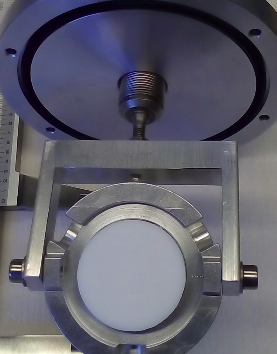}
                \caption{The PTFE sample placed in the movable sample holder.}
                \label{holder}  
                \end{center}    
      \end{figure}

The PTFE sample was cleaned with pure acetone in an ultrasonic bath and then kept under vacuum for one day. This cleaning process reduces the amount of oils present at the surface of the material. The sample can be seen in \autoref{holder}.

The two PMMA samples were cut from 1) the acrylic batch used to form the DEAP cryostat, denoted `RPT acrylic', and 2) the acrylic batch from which the DEAP light guides are made, denoted `Spartech acrylic'. The sample surfaces were polished to optical quality and then cleaned by ultrasonic cleaning in ultra-pure water (UPW) with Alconox detergent, followed by ultrasonic cleaning in UPW. The fluorescence response was measured, then the samples were cleaned again in the following way: ultrasonic bath in distilled water with industrial detergent, then in distilled water, then air plasma cleaned.

\begin{figure}[htp] 
                \begin{center}
              \includegraphics[scale=0.35]{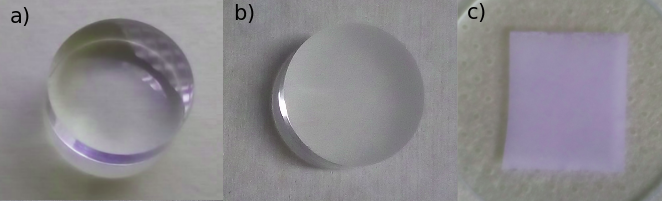}
                \caption{a) RPT acrylic sample before sanding. b) RPT acrylic sample after sanding. c) TPB reference sample.}
                \label{TPBsample}  
                \end{center}    
\end{figure}

The acrylic on the inside of the DEAP cryostat has a rough surface finish. This could change the fluorescence response for a number of reasons. Among them are: i) residue from the sandpaper could fluoresce and ii) the optical path of light is different when reflecting off a specular versus a rough surface, which could influence our measurement, especially if the light emission is not isotropic. One side of each acrylic sample was therefore sanded using the same sandpaper used on the DEAP cryostat. After sanding, the samples were sonicated in isopropanol and distilled water for five minutes at room temperature. Then, they were dried with a nitrogen gun\footnote{The DEAP cryostat acrylic inner surface was washed with ultra-pure water after resurfacing, and was kept under a nitrogen atmosphere.}.
Pictures of the acrylic samples are shown in \autoref{TPBsample}.

The TPB reference sample (\autoref{TPBsample}) was vacuum evaporated on glass. The film thickness was measured in four locations with a profilometer as \SI{1.2 \pm 0.2}{\micro\meter} (mean of the four measurements and their standard deviation).

We use an aluminum disk as a blank reference sample, because pure aluminum does not fluoresce. The sample was sonicated in distilled water with industrial detergent and then in distilled water. It was then air plasma cleaned. 

We used a borosilicate glass microscope slide to characterize background fluorescence from optical components, especially the lens, which observes the sample. Since the lens was not plasma cleaned, the glass slide was not plasma cleaned either. It was sonicated in acetone and isopropanol and dried with a nitrogen gun.


\section{Setup}
\label{sec:setup}

\begin{figure}[htp] 
                \begin{center}
                \includegraphics[scale=0.5]{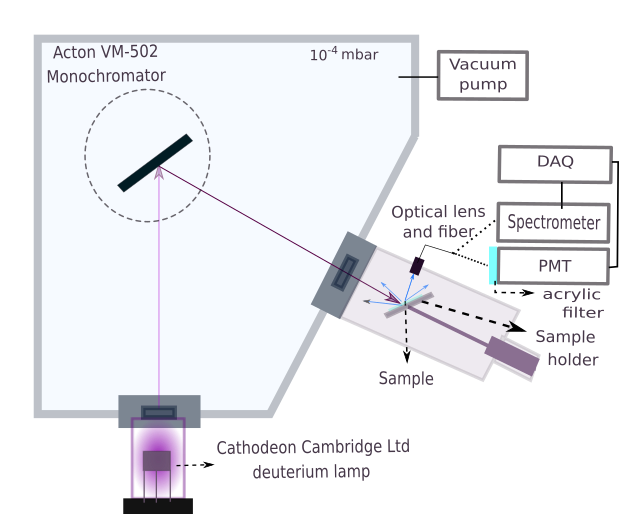}
                \caption{Schematic drawing of the fluorescence measurement setup. Light from the deuterium lamp is directed through an entrance slit toward the monochromator, which reflects the desired wavelength toward an adjustable exit slit and onto the sample. The sample is placed in a movable sample holder so that the incident angle can be varied. An optical fiber collects light reflected and re-emitted from the sample and guides it to the light detectors. The whole system is under vacuum so that the VUV light is not attenuated.}
                \label{setup}  
                \end{center}    
        \end{figure}

The setup is schematically shown in \autoref{setup}. UV light is provided by a Cathodeon deuterium lamp with a MgF$_{2}$ window. The light enters a VM-502 vacuum monochromator with a VUV reflecting 1200 lines/mm grid grating. Its FWHM resolution, at the entrance and exit slit widths used here, is approximately \SI{20}{\nano\meter}. Photons of the wavelength selected on the monochromator are guided through the exit slit and onto the sample. The sample is installed in a tube sealed against the monochromator. The pressure inside the monochromator and the sample tube was always lower than \SI{5.4e-4}{\milli\bar}. At this pressure, more than 99\% of the VUV light reaches the sample. A quartz lens is installed on the sample tube and pointed at the surface of the sample.

The sample holder, shown in \autoref{holder}, was built to accommodate the different dimensions of the samples in such a way that the illuminated area on the surface of each sample is at the same position relative to the incoming beam and the lens. The sample holder head is adjustable such that the incidence angle can be varied, and the area backing on the sample was painted black. The lens is coupled to an optical fibre which guides the photons to one of two exchangeable photon detectors: a QE65000 Ocean Optics spectrometer, or a PMT with an S20 cathode and a MgF$_{2}$ window. The PMT was operated in photon-counting mode. A thin acrylic window was used as a filter in front of the PMT to absorb light from reflections of excitation photon. 

\autoref{LArTPBresponse}c-d shows the photon emission, transmission, and detection efficiencies of the components in the setup as a function of wavelength.

The SpectraSuite software from Ocean Optics was used to acquire the data from the spectrometer at a resolution of approximately \SI{0.8}{\nano\meter} \citep{Oceanspectrometer}. The data acquisition for the PMT consisted of a  Canberra 2128 constant fraction discriminator connected to a Wenzel Elektronik counter which outputs the average rate of pulses above threshold.

\section{Procedures}
\label{sec:Proc}

We used the spectrometer to measure the fluorescence spectra, and the PMT to measure the wavelength-integrated fluorescence intensity. Measurements were done for excitation wavelengths of \SIlist{130; 150; 160; 170; 180; 210;250}{\nano\meter}, and under incidence angles of 39.5$^\circ$ and 12.5$^\circ$.

\subsection{Sample measurements}
\label{subsec:spec}

Measurements were done with the monochromator slits \SI{5}{\milli\meter} wide open.

For the wavelength-resolved measurements, the spectrometer was set up to record 10 spectra each with \SI{10}{\second} exposure time. The 10 spectra were averaged to obtain the final spectrum.

For the wavelength-integrated measurements, the optical fiber was coupled to the PMT.
The PMT rate was measured for \SI{50}{\second} ten times in a row and then averaged.

\subsection{Stability and calibration measurements}

The intensity of the deuterium lamp was stable at the 1\% level approximately \SI{40}{\minute} after turning it on. The PMT dark count rate was stable to within 10\% approximately \SI{60}{\minute} after turning on the bias voltage. Hence, wavelength-integrated measurements were performed at least \SI{60}{\minute}, and wavelength-resolved measurements at least \SI{40}{\minute} after the lamp and PMT were turned on. During this time, the exit slit on the monochromator was kept shut to prevent the samples from degrading\footnote{The WLSE of TPB is known to degrade with time when exposed to UV light~\citep{TPBdegradation}.}. 
The wavelength-dependent response of the spectrometer was calibrated with a tungsten halogen calibration lamp.

Whenever a new sample was installed in the sample holder, we evacuated the setup and, in addition to the normal measurement with open slits, recorded data while the emission slit on the monochromator was closed. This data was taken as a measure for the background light level and dark count rate, $R_\text{BDN}$.

Monochromators often transmit light at wavelengths outside the selected bandwidth \citep{straylight}. To check for this kind of stray light, we set the monochromator to pass either \SI{130}{\nano\meter} or \SI{160}{\nano\meter} light but did not evacuate the chamber. The oxygen in the air acts as a VUV filter, absorbing all the light below \SI{180}{\nano\meter}, so that only stray light above \SI{180}{\nano\meter} can reach the sample. The signal seen in this configuration is composed of: i) stray light - above the transmittance of the acrylic filter - reflected by the samples, ii) photoluminescence induced in the sample by stray light, and iii) the background light level and dark count rate as seen with the emission slit closed ($R_\text{BDN}$). Since contributions i) and ii) cannot be separated, we use $R_\text{stray}$ for their sum.

The low-intensity fluorescence of optical glasses is a known source of background in optical devices \citep{Glassfluor,Glassfluor2}. We performed two measurements to determine the possible stray fluorescence from the glass lens: i) we measured a borosilicate glass sample to obtain a rough estimate of the fluorescence signal from glass, and ii) we measured an aluminum blank, which does not fluoresce itself but reflects incident light onto the lens, possibly inducing the lens to fluoresce.

\section{Analysis}
\label{sec:analysis}
\begin{figure}[htp] 
                \begin{center} \includegraphics[scale=0.245]{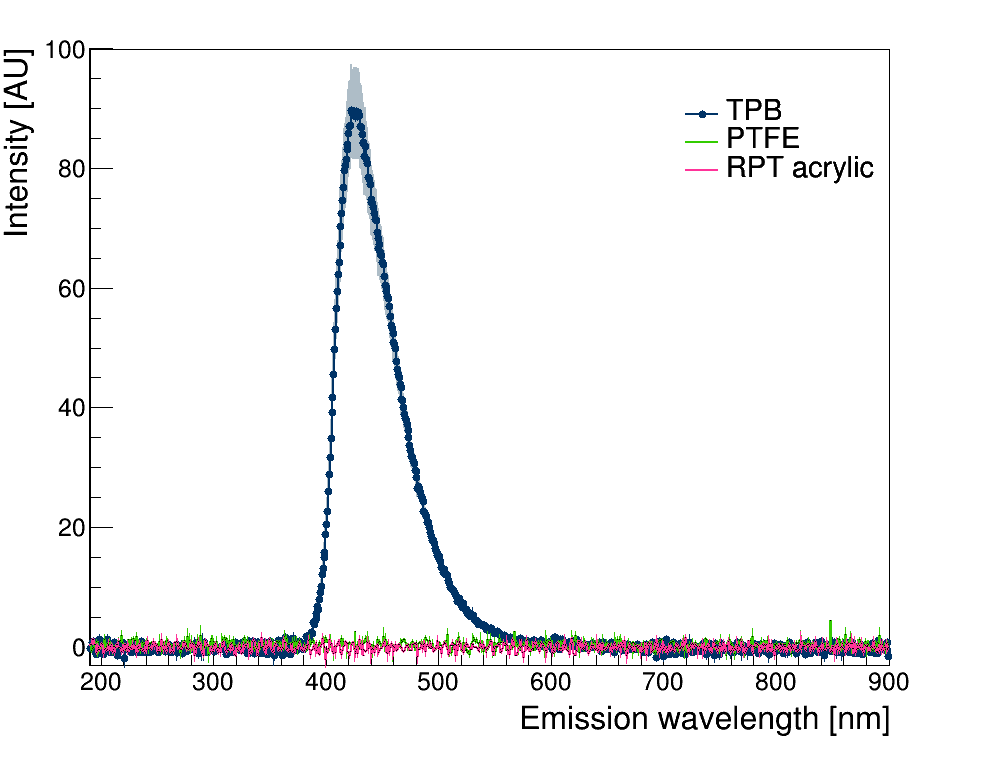}
                \caption{Emission spectrum of TPB and no signal above noise from PTFE and RPT acrylic. All the samples were measured with \SI{160}{\nm} excitation light at an incident angle of 39.5$^{\circ}$. The error bars are dominated by the wavelength-dependent systematic uncertainty of the response function.}
                \label{spectra}  
                \end{center}    
\end{figure}

The spectra from the spectrometer were background corrected with the background spectra taken from the closed-slit measurement. The TPB spectrum was corrected for the response function of the spectrometer. The final spectra for an excitation wavelength of \SI{160}{\nano\meter} are shown in \autoref{spectra}. The error bars are dominated by the wavelength-dependent systematic uncertainty of the response function.

The PMT rates from the samples were corrected for the background light level and dark rate ($R_\text{BDN}$), pulse pile-up, and dead time of the data acquisition system, resulting in $R_\text{sample}$. Typical values for $R_\text{BDN}$ are listed in \autoref{tab:bg}. Pile-up was less than 0.5\% of the signal intensity and the dead time of the counter was at most 14\% for TPB excited with \SI{160}{\nano\meter} light, where the intensity of the deuterium lamp peaks. For other excitation wavelengths the dead time was less than 3\%, and for the plastic samples the dead time was less than 0.03\%. 

After these corrections, the mean rates of the measurements of the plastic samples were divided by the mean rate of the TPB sample measured at the same angle and with the same excitation wavelength, resulting in the relative signal intensity. 


\section{Results and discussion}
\label{sec:results}

The sensitivity of our setup is best at the intensity peak of the deuterium lamp (\SI{160}{\nano\meter}). At this excitation wavelength, the spectrum of TPB is shown in \autoref{spectra}. Also shown are the measurements of PTFE and RPT acrylic done under the same conditions. If PTFE or acrylic display photoluminescence in the visible regime when excited by VUV light, the light yield is below the sensitivity of the spectrometer. The integrated signal to noise ratio is approximately 1\% in the wavelength range of the emission of TPB. Hence, any fluorescence signal in this range must be below the 1\% level.

The measurements with the PMT have a better signal to noise ratio, since the PMT integrates the signal over a range of wavelengths. While some light was observed from the samples, there are several sources of background. Typical rates are listed in \autoref{tab:bg} as an example.

\begin{table}[htpb]
    \centering
    \caption{Background and signal rates observed with the PMT for samples measured with \SI{160}{\nm} excitation light at 39.5$^{\circ}$ incidence angle. $R_\text{stray}$, and the sample rates of aluminum and glass are used to estimate the background light level in the signal observed from the plastic and TPB samples.}
    \begin{tabular}{l|l|l|l}
\textbf{Sample} & \multicolumn{3}{c}{\textbf{Contribution}}\\
    &  $R_\text{BDN}$ [Hz] & $R_\text{stray}$[Hz] & $R_\text{sample}$ [Hz] \\
\hline 
aluminum & 14 & 4 & 52 \\
glass & 18 & 0 & 374 \\
TPB & 23 & 94 & 355462 \\
PMMA (RPT: smooth) & 16 & 0 & 24 \\
PMMA (RPT: rough) & 16 & 1 & 58 \\
PTFE & 25 & 39 & 80 \\

    \end{tabular}
    \label{tab:bg}
\end{table}





In the measurement of the aluminum disk, $R_\text{sample}$ was higher than $R_\text{stray}$, indicating that the excitation light reflected from the aluminum induces fluorescence somewhere in the setup, most likely in the lens. This is also supported by the fact that the $R_\text{sample}$ of the glass slide was more intense than that from any of the plastic samples, indicating that the setup is sensitive to fluorescence from glasses. Even though the measurements of the glass and aluminum samples indicate non-negligible levels of background fluorescence, we cannot subtract these values from the $R_\text{sample}$ of the plastic samples because each sample reflects the excitation light differently, inducing fluorescence at different intensities in the lens. Furthermore, the $R_\text{sample}$ rates from plastic and aluminum are at the same level, so that this background cannot be subtracted with any confidence. It was also not possible to assess the level of stray light from the monochromator at excitation wavelengths above approximately \SI{180}{\nano\meter}, since air can no longer be used as a filter in that region. Thus, there is no measurement of this rate that could be subtracted from the sample rate. 

\begin{figure}[htp] 
\begin{center}
    \includegraphics[scale=0.225]{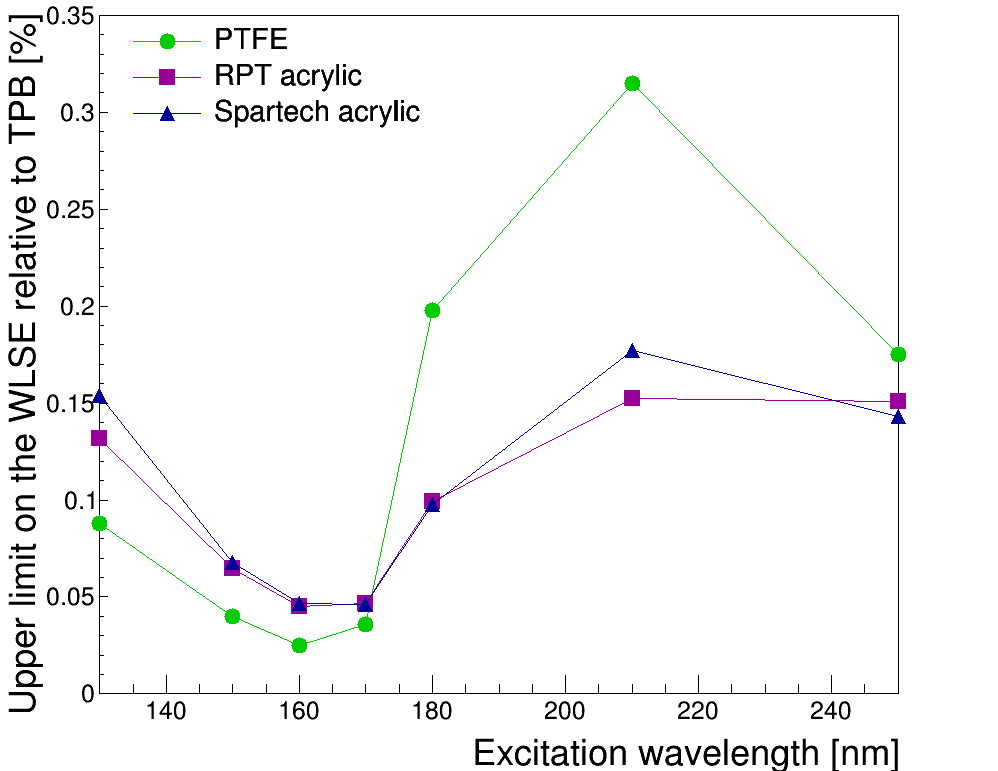}
    \caption{Upper limit on the wavelength shifting efficiency of PTFE ($\smallblackcircle$) , RPT acrylic ($\smallblacksquare$), and Spartech acrylic ($\smallblacktriangleup$) relative to TPB for different excitation wavelengths. Lines are shown only to guide the eyes.}
    \label{limits}     
\end{center}    
\end{figure}

In light of these backgrounds, we only derive upper limits on the wavelength shifting efficiency of PTFE and acrylic relative to TPB. The limits shown in \autoref{limits} correspond to the mean of the measured light level plus 1.645 times the uncertainty. Since much of the measured rate is from backgrounds, the limits are conservative. 


To derive the upper limit on the relative WLSE of PTFE, we used the results from the measurements done at 39.5$^{\circ}$, which had higher rates than the measurements done at 12.5$^{\circ}$. The difference could be due to: i) the angle dependence of reflected light, which is especially important for PTFE, since it is a reflector, and/or ii) differences in the angular distribution of light emitted by PTFE and TPB.

The measurements of the acrylic samples at different angles of incidence agreed within errors. For the smooth samples, the rate was higher before plasma cleaning. This could be due to fluorescent surface contaminants, such as oils, present on the samples before plasma cleaning, and points to the importance of surface treatment. However, the rate decrease could also have been due to damage to the organic bonds of PMMA from energetic-particle and/or VUV-photons created in the plasma cleaning process. We have not observed any decrease in the rate from the acrylic samples as a function of illumination time. That is, VUV light did not seem to degrade the samples.  Furthermore, if the organic bonds were superficially damaged, the sanding of the sample would expose a non-degraded surface to the excitation light. 

The rough samples did have a slightly higher rate than the smooth ones. However, this could also be due to: i) differences in the paths of excitation light after reaching the surface, ii) less angle-dependence of the light emission distribution, iii) residue from the sandpaper. To rule out point (iii) we measured the sandpaper and observed no fluorescence signal from it.

We used the more conservative results from the rough acrylic samples to derive the upper limit on the relative WLSE of acrylic shown in \autoref{limits}. 


The results shown in \autoref{limits} are obtained under the assumption that the emission spectra of the samples are similar to that of TPB. Since at least DEAP PMTs are most sensitive in this region, the results are still relevant. If acrylic or PTFE are found to display photoluminescence with spectra different from TPB, the results must be corrected for the wavelength-dependent sensitivity of the setup.

To transfer results to experiments that do not use TPB, the thickness-dependent WLSE of TPB can be found in \citep{TPBGehman2, TPBDavis}.


\section{Conclusions}
\label{sec:conclusions}
We investigated the photoluminescence response of acrylic and PTFE to excitation light between \SIrange{130}{250}{\nano\meter} and emission between  \SIrange{400}{550}{\nano\meter}  with a sensitivity relevant to rare-event searches using noble gas targets. This is the first time the photoluminescence of these plastics was investigated for such low excitation wavelengths.

 The sensitivity to photoluminescence light achieved is at the level of 10$^{-3}$ of signal strength from a \SI{1.2}{\micro\meter} thick vacuum-evaporated TPB film. The limiting factors of the sensitivity were i) stray light from the monochromator and ii) fluorescence of the optical lens, induced by excitation light reflected by the samples. We found that the cleaning procedure of the plastics may play an important role in the level of photoluminescence. 
 
 Within this sensitivity, no signal that can clearly be attributed to fluorescence from the plastics was observed. We place conservative upper limits on the WLSE of the plastics at less than 0.35\% the WLSE of TPB. We therefore find it unlikely that photoluminescence from room-temperature DEAP PMMA and LUX PTFE has a measurable effect in these detectors.

There are some indications that temperature plays a role in the fluorescence response of materials \citep{Veloce:2015slj,PENdiffT}. In the future, we plan to repeat these measurements on samples cooled to LAr and LXe temperatures.

These results are especially important for the development of future rare-event search experiments based on liquid scintillators that emit VUV light \citep{DS20K, Darwin}. They may also be relevant for future experiments that will use acrylic vessels, such as JUNO \citep{acrylJUNO}, since part of the non-visible Cherenkov light produced in acrylic could be wavelength shifted into the visible, and thus detectable, regime. The results can also be of interest for the current background analysis of experiments that use acrylic or PTFE surrounding the target, such as DEAP \citep{DEAPconst}, LUX \citep{LUX}, DarkSide \citep{DarkSide}, Xenon \citep{xenonT}, ArDM \citep{ArDM}, and SNO+ \citep{SNOplus}.

\section*{Acknowledgements}
We would like to thank Marcin Ku{\'z}niak from Carleton University for providing the acrylic samples used in DEAP and Francisco Neves from Coimbra University for providing the PTFE samples from LUX. We also thank Jochen Wieser from Excitech GmbH for the plasma cleaning of the samples and helpful discussions.

\label{Bibliography}
\bibliographystyle{tp_unsrt_doi}
\bibliography{Bibliography.bib}

\end{document}